\documentclass[preprint,onecolumn,nofootinbib]{revtex4}
\pdfoutput=1
\usepackage[colorlinks=true,linkcolor=blue,urlcolor=blue,filecolor=black,citecolor=red,pdfstartview=FitV,pdftitle={},pdfsubject={},pdfkeywords={},pdfpagemode=None,bookmarksopen=true]{hyperref}
\usepackage{graphicx}%Include figure files
\usepackage{amsmath}
\usepackage{amsfonts}
\usepackage{amssymb,ulem}
\usepackage{color}%
\usepackage{tikz}
\usepackage{dcolumn}% Align table columns on decimal point
\usepackage{xcolor}

\begin{document}
\title{Mixed State Entanglement and Thermal Phase Transitions}
\author{Peng Liu $^{1}$}
\email{phylp@email.jnu.edu.cn}
\author{Jian-Pin Wu $^{2}$}
\email{jianpinwu@yzu.edu.cn} \affiliation{
  $^1$ Department of Physics and Siyuan Laboratory, Jinan University, Guangzhou 510632, China\\
  $^2$ Center for Gravitation and Cosmology, College of Physical
  Science and Technology, Yangzhou University, Yangzhou 225009,
  China
}

\begin{abstract}

  We study the relationship between mixed state entanglement and thermal phase transitions. As a typical example, we compute the holographic entanglement entropy (HEE), holographic mutual information (MI) and the holographic entanglement of purification (EoP) over the superconductivity phase transition. We find that HEE, MI and EoP can all diagnose the superconducting phase transition. They are continuous at the critical point, but their first derivative with respect to temperature is discontinuous. MI decreases with increasing temperature and exhibits a convex behavior, while HEE increases with increasing temperature and exhibits a concave behavior. However, EoP can exhibit either the same or the opposite behavior as MI, depending on the size of the specific configuration. These results show that EoP captures more abundant information than HEE and MI.
  We also provide a new algorithm to compute the EoP for general configurations.

\end{abstract}
\maketitle
\tableofcontents

\section{Introduction}
\label{sec:introduction}

Quantum entanglement is the main property that distinguishes quantum systems from classical systems. Recently, quantum entanglement has become a hot topic in the fields of holographic gravity, condensed matter theory, quantum information and so on. Many quantum entanglement measure have been found capable of diagnosing the quantum phase transition of strong correlation systems and the topological quantum phase transitions, and playing a key role in the emergence of spacetime \cite{Osterloh:2002na,Amico:2007ag,Wen:2006topo,Kitaev:2006topo,Ryu:2006bv,Hubeny:2007xt,Lewkowycz:2013nqa,Dong:2016hjy}.

There are many different types of quantum entanglement measures, such as entanglement entropy (EE), mutual information (MI), R\'enyi entanglement entropy, negativity, and so on. Among these quantum entanglement measures, EE has been widely accepted as a good measure for pure state entanglement. However, EE is unsuitable for measuring the entanglement of mixed state, which is way more common than pure state. Many new entanglement measures have been proposed to measure mixed state entanglement, such as the entanglement of purification (EoP), non-negativity, and the entanglement of formation \cite{vidal:2002,Horodecki:2009review}. However, entanglement measures are extremely difficult to calculate.

Gauge/gravity duality provides a powerful tool for studying strongly correlated systems, and it relates entanglement related physical quantities to geometric objects in dual gravity systems. The holographic entanglement entropy (HEE) associates the EE of a subregion on the boundary with the area of the minimum surface in the dual gravity system \cite{Ryu:2006bv}. HEE has been shown capable of diagnosing quantum phase transitions and thermodynamic phase transitions \cite{Nishioka:2006gr,Klebanov:2007ws,Pakman:2008ui,Zhang:2016rcm,Zeng:2016fsb,Ling:2015dma,Ling:2016wyr,Ling:2016dck,Kuang:2014kha,Guo:2019vni,Mahapatra:2019uql}. Recently, the R\'enyi entropy has been proposed to be proportional to the minimal area of cosmic branes \cite{Dong:2016fnf}. Moreover, the butterfly effect that reflects the dynamic properties of quantum systems, has been extensively studied in holographic theory \cite{Shenker:2013pqa,Sekino:2008he,Maldacena:2015waa,Donos:2012js,Blake:2016wvh,Blake:2016sud,Ling:2016ibq,Ling:2016wuy,Wu:2017mdl,Liu:2019npm}. In addition, holographic duality of quantum complexity, a new information-related quantity from the EE, was also proposed \cite{Brown:2015lvg,Brown:2015bva,Chapman:2016hwi,Ling:2018xpc,Chen:2018mcc,Yang:2019gce,Ling:2019ien,Zhou:2019xzc,Zhou:2019jlh,Chakrabortty:2020ptb}. More recently, the EoP was associated with the area of the minimum cross-section of the entanglement wedge \cite{Takayanagi:2017knl,Nguyen:2017yqw}. The geometric prescription of EoP provides a novel and powerful tool for studying the mixed state entanglement \cite{Huang:2019zph,Fu:2020oep,Gong:2020pse,Liu:2019npm,Liu:2019qje,Lala:2020lcp,Bao:2018gck,Umemoto:2018jpc,Yang:2018gfq}.

At present, HEE has been widely studied over many different holographic phase transition models, but the research on mixed state entanglement - MI and EoP, are still missing. For this purpose, we study the properties of HEE, MI and EoP, in holographic superconductivity model. We focus on the relationship between these information-related physical quantities and phase transitions, and pay special attention to the difference and relationship between mixed state entanglement measures and HEE.

We organize this paper as follows: we introduce the holographic superconductivity model in Sec. \ref{sec:phase_transition}, entanglement measures (HEE, MI, EoP) and their holographic duality in Sec. \ref{subsec:info}. We discuss the properties of HEE (\ref{sec:hee}), MI (\ref{sec:mi}) and EoP (\ref{sec:eop_phenomena}) systematically. Finally, we summarize in Sec. \ref{sec:discuss}.

\section{Holographic superconductivity phase transition and holographic information-related quantities}\label{sec:alg}

First, we introduce the HEE, MI and EoP and its holographic dual. After that, we elaborate on the new algorithms to compute the minimum surfaces and minimum cross-sections.
\subsection{Holographic superconductivity phase transition}\label{sec:phase_transition}

A thermal phase transition occurs at a finite temperature, that usually is accompanied by a symmetry breaking and the emergence of an order parameter. A prominent example of the holographic thermal phase transition is the superconductivity phase transition model. The action of the holographic superconductor is \cite{Hartnoll:2008vx} (see also \cite{Cai:2015cya} for a recent review),
\begin{equation}
  \label{eq:holoaction}
  S = \int d ^ { 4 } x \sqrt { - g } \left( R + \frac { 6 } { L ^ { 2 } } - \frac { 1 } { 2 } F _ { \mu \nu } F ^ { \mu \nu } - | \nabla \Psi - i q A \Psi | ^ { 2 } - m ^ { 2 } \left| \Psi ^ { 2 } \right| \right),
\end{equation}
where $L$ is the AdS length scale, $A$ and $F=dA$ is the gauge field and the corresponding field strength. $\Psi$ is the complex scalar field representing the superconductivity condensation, which we write as $\Psi = e^{i\xi} \psi$ with $\psi$ a real scalar field and $\xi$ the St\"uckelberg field. We fix the gauge by setting $\xi=0$, and the corresponding equations of motion read,
\begin{equation}
  \label{eq:acteom}
  \begin{aligned}
    R _ { \mu \nu } + g _ { \mu \nu } \left(  3  + \psi ^ { 2 } \right) - \left( F _ { \mu \lambda } F _ { \nu } {}^ { \lambda } - \frac { 1 } { 4 } g _ { \mu \nu } F ^ { 2 } \right) - \partial _ { \mu } \psi \partial _ { \nu } \psi - q ^ { 2 } \psi ^ { 2 } A _ { \mu } A _ { \nu } & = 0, \\
    \nabla _ { \mu } F ^ { \mu } {}_ { \nu }  - q ^ { 2 } \psi ^ { 2 } A _ { \nu } = 0,\quad \left( \nabla ^ { 2 } - q ^ { 2 } A ^ { 2 } + 2 \right) \psi                                                                                                                                 & = 0.
  \end{aligned}
\end{equation}
We solve them with ansatz,
\begin{equation}
  \label{eq:ansatz}
  \begin{aligned}
    d s ^ { 2 } & = \frac { 1 } { z ^ { 2 } } \left[ - ( 1 - z ) p ( z ) U d t ^ { 2 } + \frac { d z ^ { 2 } } { ( 1 - z ) p ( z ) U } + V d x ^ { 2 } + V d y ^ { 2 } \right], \\
    A           & = \mu ( 1 - z ) a dt,
  \end{aligned}
\end{equation}
where $\mu$ is the chemical potential of the gauge field $A$ and $p(z)\equiv 1 + z + z ^ { 2 } - \mu ^ { 2 } z ^ { 3 } / 2$. The $z$ is the radial axis, and $z=0,\,1$ represents the AdS boundary and the horizon, respectively. The quantities $U, \,V,\, a$ and $\psi$ are all functions of $z$, that can be obtained by solving the equations of motion \eqref{eq:acteom}. The system has a simple solution with $a=U=V=1,\,\psi = 0$, where the system goes back to the AdS-RN black brane.

\begin{figure}[]
  \centering
  \includegraphics[width =0.6\textwidth]{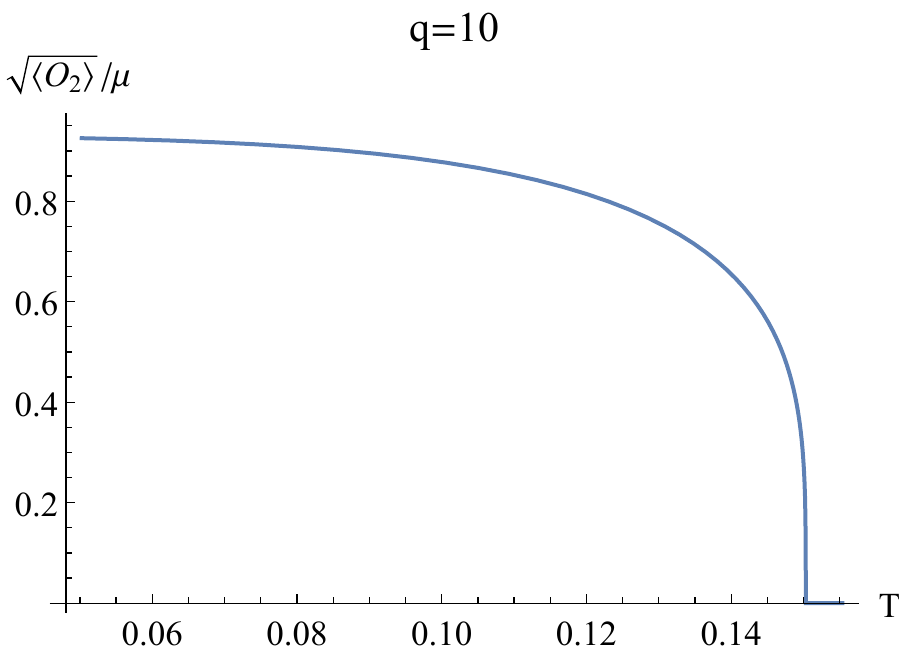}
  \caption{The condensation at $q=10$. The critical temperature $T_c=0.150257$, below which the system is in the superconducting phase.}
  \label{fig:condensation}
\end{figure}

The Hawking temperature is $\tilde T = \frac{6-\mu^2}{8\pi}$. The system has a scaling symmetry,
\begin{equation}
  \label{eq:scaling}
  \mu\to \alpha\mu,\quad \tilde T\to \alpha \tilde T,\quad V\to \alpha^{ -2} V.
\end{equation}
In this paper, we adopt $\mu$ as the scaling unit\footnote{This is equivalent to choosing the grand canonical ensemble to describe the system.}, and the dimensionless Hawking temperature $T = \tilde T/\mu$. For concreteness, we fix $L=1,\,m^2=-2$ and $q=10$ where the critical temperature $T_c = 0.150257$. We show the condensation $\sqrt{\langle O_2 \rangle}/\mu$ vs $T$ in Fig. \ref{fig:condensation}.

\subsection{Holographic information-related quantities}\label{subsec:info}

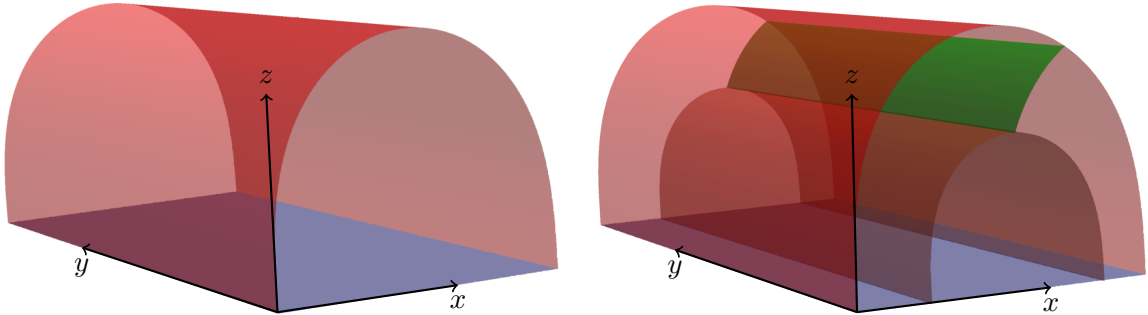
\begin{figure}
  \begin{tikzpicture}[scale=1]
    \node [above right] at (0,0) {\includegraphics[width=7.5cm]{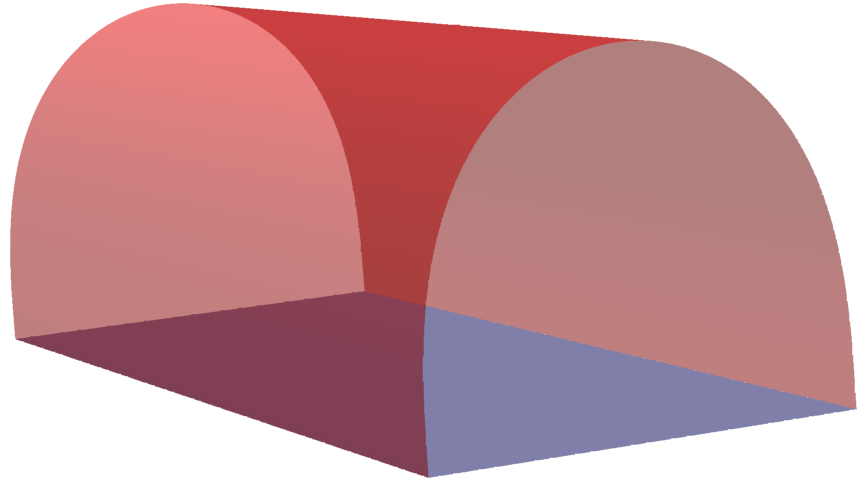}};
    \draw [right,->,thick] (3.85, 0.22) -- (6.25, 0.58) node[below] {$x$};
    \draw [right,->,thick] (3.85, 0.22) -- (1.25, 1.08) node[below] {$y$};
    \draw [right,->,thick] (3.85, 0.22) -- (3.7, 3.125) node[above] {$z$};
  \end{tikzpicture}
  \begin{tikzpicture}[scale=1]
    \node [above right] at (0,0) {\includegraphics[width=7.5cm]{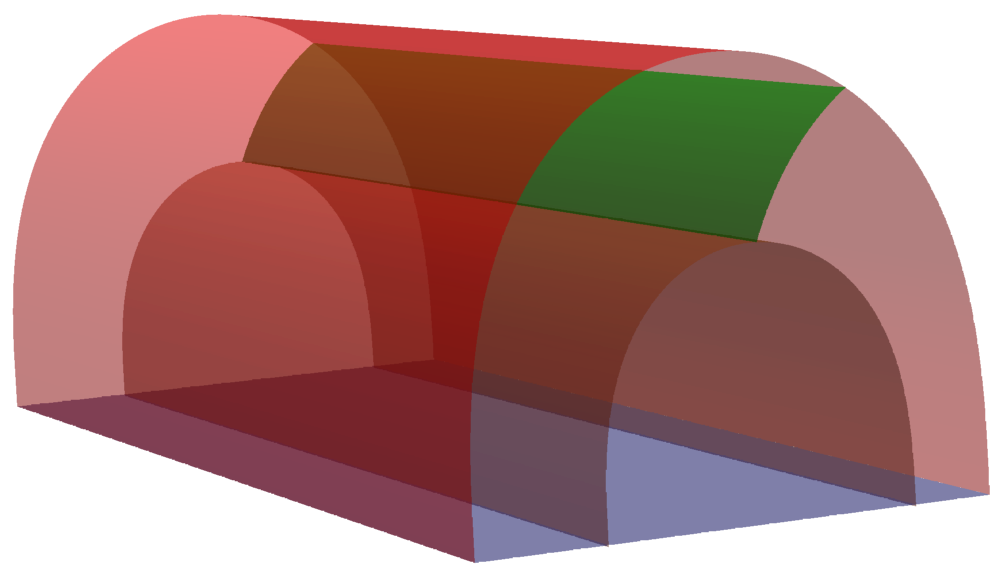}};
    \draw [right,->,thick] (3.67, 0.22) -- (6.25, 0.55) node[below] {$x$};
    \draw [right,->,thick] (3.67, 0.22) -- (1.25, 1.05) node[below] {$y$};
    \draw [right,->,thick] (3.67, 0.22) -- (3.6, 3.125) node[above] {$z$};
  \end{tikzpicture}

  \caption{The left plot: The minimum surface for a given width $w$. The right plot: The minimum cross-section (green surface) of the entanglement wedge.}
  \label{msd1}
\end{figure}

One of the most striking features of quantum mechanics is the entanglement. The most famous measure of entanglement is EE, which measures the entanglement between a subsystem and its complement. Specifically, the EE $S_A$ between $A$ and $B$ in $A\cup B$ is defined as von Newmann entropy in terms of the reduced density matrix $\rho_A$,
\begin{equation}\label{ee-von}
  S_{A} (|\psi\rangle) = - \text{Tr}\left[ \rho_{A} \log \rho_{A} \right],\quad \rho_{A} = \text{Tr}_{B} \left(|\psi\rangle\langle\psi|\right).
\end{equation}
For pure states we will find that $S_{A} = S_{B}$ \cite{Chuang:2002book}. In holographic duality theory, the HEE was related to the area of the minimum surface in dual gravity systems \cite{Ryu:2006bv} (see the left plot of Fig. \ref{msd1}).

EE has been widely accepted as a good measure of pure state entanglement. However, EE is not a good measure of mixed state entanglement. For instance, $A$ and $B$ in a system formed by the direct product of density matrix $\rho_A$ and $\rho_B$ does not entangle with each other, but can have non-zero EE. The reason is that EE not only takes into account the quantum entanglement, but also the classical correlation. Many new measures of the mixed state entanglement have been proposed \cite{vidal:2002,Horodecki:2009review}, among which the most direct measure of mixed state entanglement is the mutual information (MI).

For separate $A\cup B$, the MI is defined as
\begin{equation}\label{mi:def}
  I\left(A,B\right) := S\left(A\right) + S\left(B\right) - S\left(A\cup B\right),
\end{equation}
which measures the entanglement between $A$ and $B$. It is easy to verify that $ I\left(A,B\right) =0 $ when $\rho_{AB} = \rho_{A} \otimes \rho_{B} $. Therefore, MI exhibits the important property that the direct product state has zero entanglement. However, MI is also not a perfect measure for mixed state entanglement. Since MI is defined by EE in essence, \cite{Huang:2019zph} points out that the properties of MI in some cases are completely dominated by EE or even thermal entropy. This shows that we need to resort to other mixed state entanglement measures.

The EoP, which involves the purification of mixed states, is a new mixed state entanglement measure that is currently being extensively studied. EoP has been shown to satisfy several important inequalities \cite{Terhal:2002}. Therefore, a reliable holographic dual must also satisfy these inequalities \cite{Takayanagi:2017knl,Bao:2017nhh}. Takayanagi proposed a holographic dual of the EoP $ E_{W}\left(\rho_{AB}\right) $ as the area of the minimum cross-section $ \Sigma_{AB} $ in connected entanglement wedge \cite{Takayanagi:2017knl}, {\it i.e.}, the configurations with non-zero MI (see the right plot in Fig. \ref{msd1}),
\begin{equation}\label{heop:def}
  E_{W}\left(\rho_{AB}\right) = \min_{\Sigma_{AB}} \left( \frac{\text{Area} \left(\Sigma_{AB}\right)}{4G_{N}}\right) .
\end{equation}
It is worth noting that the EoP, i.e., the minimum cross-section can only exist in the connected entanglement wedge. For disconnected cases where MI vanishes, the EoP also vanishes.

The EoP lives in the entanglement wedge that relates to the minimum surfaces. Therefore, we provide new algorithms to obtain the minimum surfaces and the EoP. HEE, MI, and EoP all have the same scaling dimension, and we divide them by $\mu$ to obtain the dimensionless quantities.

\subsection{Computations of holographic EoP}\label{eopcomp}

For the convenience of numerics, we study the EoP of infinite strips in homogeneous background. For a generic homogeneous background
\begin{equation}\label{genbg}
  ds^{2} = {g_{tt}} dt^2 + g_{zz}dz^2 + g_{xx}dx^2 + g_{yy} dy^2,
\end{equation}
with $z=0$ denoting the asymptotic AdS boundary, the left plot in Fig. \ref{fig:cartoon4eop} is the cartoon of the minimum surface for an infinite strip along $y$-axis. The homogeneity requires that all metric components $g_{\mu\nu}$ are functions of $z$ only.

In previous work \cite{Liu:2019qje}, we used NDSolve with Mathematica to develop an algorithm to solve the minimum surface and the corresponding asymmetric EoP. We adopted arc-length parameter and took full advantage of homogeneity to accelerate the solution of EoP, and applied this algorithm to calculate asymmetric EoP in AdS$_4$ space-time and AdS-RN black hole systems. However, we encountered some limitations with this algorithm. First of all, choosing the arc-length parameter will make it difficult to solve the minimum surface in the asymptotic AdS region. Secondly, NDSolve method fails easily in the near horizon region due to the coordinate singularity. As a result, this algorithm can only offer reliable numerical EoP results in a relatively narrow range of parameters. In this paper, we propose new algorithms to calculate the minimum surface and asymmetric EoP, that will render the numerical computation much more stable and reliable.

\subsubsection{The minimum surface}

\begin{figure}
  \centering
  \includegraphics[width=\textwidth]{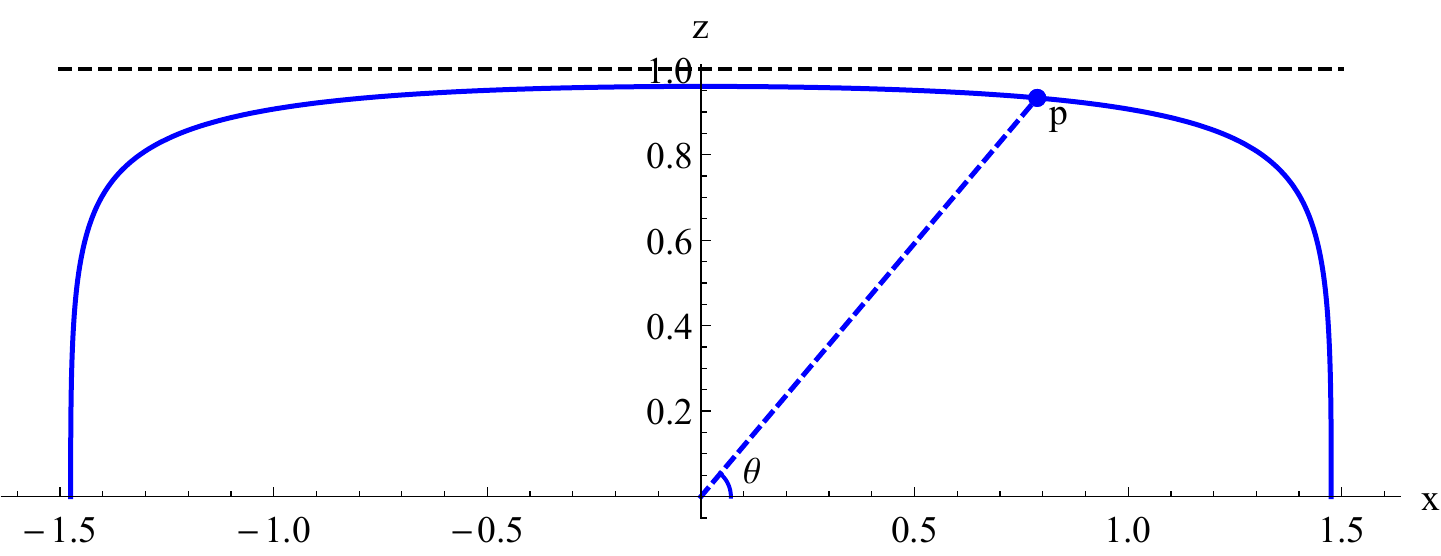}
  \caption{The angle as the parameter of the minimum surface. The horizontal black line is the horizon $(z=1)$.}
  \label{fig:angleshow}
\end{figure}

The minimum surface near the AdS boundary is perpendicular to the boundary, which renders the spatial direction $x$ an unsuitable parameter for solving the minimum surface. Ref. \cite{Umemoto:2018jpc,Ling:2019tbi} adopted the angle $\theta$ with $\tan\theta = z/x$, as the parameter of the minimum surface (see Fig. \ref{fig:cartoon4eop}). As shown in Fig. \ref{fig:angleshow}, the homogeneity of the background ensures that the minimum curve is symmetrical about the middle vertical line, which renders $\theta$, the angle between the line from the origin to the point on the curve and the $x$-axis, a good parameterization of the curve. The angle $\theta$ ranges between $[0,\pi/2]$, and the full solution on $[0,\pi]$ can be obtained by mirroring the solution on $[0,\pi/2]$ to $[\pi/2,\pi]$. We follow this method, and a surface can be parametrized as $(x(\theta),z(\theta))$ with area $A$ given by
\begin{equation}\label{eq:lag}
  A = 2\int_{0}^{\pi/2} \sqrt{x'(\theta )^2 g_{xx} g_{yy}+z'(\theta )^2 g_{yy} g_{zz}} d\theta.
\end{equation}
The resultant equations of motion read,
\begin{equation}\label{eq:eom}
  \begin{aligned}
    x'(\theta ) z'(\theta )^2 \left(\frac{g_{ xx }'}{2 g_{ xx }}+\frac{g_{ yy }'}{g_{ yy }}-\frac{g_{ zz }'}{2 g_{ zz }}\right)+\frac{x'(\theta )^3 \left(g_{ yy } g_{ xx }'+g_{ xx } g_{ yy }'\right)}{2 g_{ xx } g_{ zz }}+x''(\theta ) z'(\theta )-x'(\theta ) z''(\theta ) & =0, \\
    z(\theta) - \tan(\theta) x(\theta)                                                                                                                                                                                                                                         & =0.
  \end{aligned}
\end{equation}
where $g_{\#\#}'\equiv g_{\#\#}'(z)$.
However, it seems that the second equation in \eqref{eq:eom} can be substituted into the first equation to eliminate $x(\theta)$ or $z(\theta)$. This is feasible in principle. However, the singularity of $\tan(\pi/2)$ and possible $\tan(0)$ in the denominator will plague the numerics. Therefore, we adopt the seemingly redundant formalism in order to obtain a precise enough solution. The boundary conditions are,
\begin{equation}\label{eq:bcs}
  z(0)=0,\quad x(0)=w,\quad z'(\pi/2)=0,\quad x(\pi/2)= 0,
\end{equation}
where $w$ is the width of the infinite strip. Also, $x(\pi/2)=0$ constraint the origin $x=0$ as the middle of the minimum surface, and $z'(\pi/2)=0$ reflects the fact that the minimum surface symmetry about the middle of the minimum surface.

In order to solve \eqref{eq:eom} with first order boundary condition \eqref{eq:bcs}, we discretize the $\theta$ with finite difference method\footnote{One should choose the Gauss-Lobatto allocation for better numerical convergence. The full-order finite difference method on the Gauss-Lobatto allocation is essentially equivalent to the pseudo-spectral method using Chebyshev basis \cite{Boyd:2001}. For problems with moving endpoints, the finite difference method is more flexible.} and attack the non-linearity with the Newton-Raphson iteration method. These methods are also used in \cite{Ling:2015dma,Ling:2016wyr,Ling:2016dck} to solve numerical holographic systems. Compared with the built-in NDSolve, this method is insensitive to the coordinate singularity of the near-horizon geometry, thus improving the accuracy of the minimum surface solution. Based on the minimum surfaces, we use the Newton-Raphson iteration method again to solve the area of the minimum cross-section between the two minimum surfaces, i.e., the EoP.

\subsubsection{The EoP (minimum cross-section)}

In \cite{Liu:2019qje}, we transform the solution of EoP into a problem of solving the minimum value in two-dimensional space. In fact, the globally minimum cross-section must be orthogonal to the minimum surface at the intersections, since a global minimum must also be local minimum. This local constraint can be used to accelerate the search of the minimum cross-section, since it does not need to compute the arc length.

Given a biparty subsystem with minimum surfaces $C_1(\theta_1),\,C_2(\theta_2)$, we solve the minimum surface $C_{p_1,p_2}$ connecting $p_1 \in C_1$ and $p_2\in C_2$. We parametrize $C_{p_1,p_2}$ with $z$, then the area of $C_{p_1,p_2}$ reads,
\begin{equation}\label{eq:zpara}
  A = \int_{C_{p_1,p_2}} \sqrt{ g_{xx} g_{yy} x'(z)^2 + g_{xx} g_{zz} } dz.
\end{equation}
The resultant equation of motion becomes,
\begin{equation}\label{eq:zparaeom}
  x'(z)^3 \left(\frac{g_{ xx } g_{ yy }'}{2 g_{ yy } g_{ zz }}+\frac{g_{ xx }'}{2 g_{ zz }}\right)+x'(z) \left(\frac{g_{ xx }'}{g_{ xx }}+\frac{g_{ yy }'}{2 g_{ yy }}-\frac{g_{ zz }'}{2 g_{ zz }}\right)+x''(z) =0,
\end{equation}
with boundary condition,
\begin{equation}\label{eq:zparabcs}
  x(z(\theta_1)) = x(\theta_1),\quad x(z(\theta_2)) = x(\theta_2).
\end{equation}

\begin{figure}
  \centering
  \includegraphics[width = \textwidth]{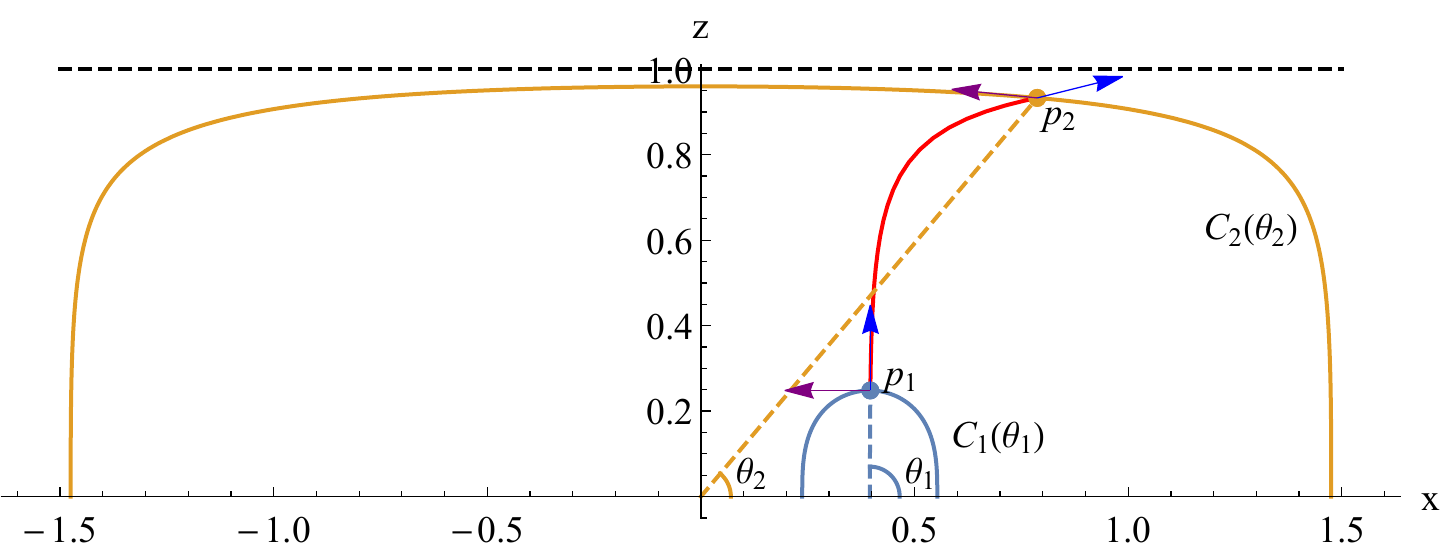}
  \caption{
  The demonstration of the EoP. The $p_1$ and $p_2$ are the intersection points of the minimum surface connecting those two minimum surfaces. The solid blue curve (parametrized with $\theta_1$) and solid orange curve (parametrized with $\theta_2$) are minimum surfaces. The thick red curve is the minimum surface connecting $p_1$ and $p_2$. The blue arrows at the $p_1$ and $p_2$ are the tangent vector $\left.\left(\frac{\partial}{\partial z}\right)^a\right|_{p_1}$ and $\left.\left(\frac{\partial}{\partial z}\right)^a\right|_{p_2}$ along the $C_{p_1,p_2}$, while the purple arrows are the tangent vectors $\left.\left(\frac{\partial}{\partial \theta_1}\right)^a\right|_{p_1}$ and $\left.\left(\frac{\partial}{\partial \theta_2}\right)^a\right|_{p_2}$ along $C_1,\,C_2$, respectively. The dark dashed horizontal line is the horizon.}
  \label{fig:cartoon4eop}
\end{figure}

We show in Fig. \ref{fig:cartoon4eop} the methods for solving the EoP.
The perpendicular conditions at the endpoints read,
\begin{equation}\label{eq:perpend}
  \left.g_{ab} \left(\frac{\partial}{\partial z}\right)^a \left(\frac{\partial}{\partial \theta_1}\right)^b\right|_{p_1} = 0, \quad \left.g_{ab} \left(\frac{\partial}{\partial z}\right)^a \left(\frac{\partial}{\partial \theta_2}\right)^b\right|_{p_2} = 0.
\end{equation}
Now, solving the EoP is to find the minimum surface ending at $(\theta_1,\theta_2)$ where \eqref{eq:perpend} is satisfied. Notice that vectors $\frac{\partial}{\partial z},\,\frac{\partial}{\partial \theta_1},\,\frac{\partial}{\partial \theta_2}$ are not normalized, for numerical stability it is better to implement the perpendicular conditions with normalized vectors as,
\begin{equation}\label{eq:perpend2}
  \begin{aligned}
    Q_1(\theta_1,\theta_2) & \equiv \left.\frac{g_{ab} \left(\frac{\partial}{\partial z}\right)^a \left(\frac{\partial}{\partial \theta_1}\right)^b}{\sqrt{g_{cd} \left(\frac{\partial}{\partial z}\right)^c \left(\frac{\partial}{\partial z}\right)^d} \sqrt{g_{mn} \left(\frac{\partial}{\partial \theta_1}\right)^m \left(\frac{\partial}{\partial \theta_1}\right)^n}}\right|_{p_1} = 0, \\
    Q_2(\theta_1,\theta_2) & \equiv \left.\frac{g_{ab} \left(\frac{\partial}{\partial z}\right)^a \left(\frac{\partial}{\partial \theta_2}\right)^b}{\sqrt{g_{cd} \left(\frac{\partial}{\partial z}\right)^c \left(\frac{\partial}{\partial z}\right)^d} \sqrt{g_{mn} \left(\frac{\partial}{\partial \theta_2}\right)^m \left(\frac{\partial}{\partial \theta_2}\right)^n}}\right|_{p_2} = 0.
  \end{aligned}
\end{equation}
Note that $Q_1$ and $Q_2$ are both functions of the $\theta_1$ and $\theta_2$. Now, the search of the EoP is equivalent to finding the minimum surface ending at $(\theta_1,\theta_2)$ where \eqref{eq:perpend2} is satisfied.

In order to find the EoP, we implement the Newton-Raphson method, that we describe in below.
\begin{enumerate}
  \item Prepare initial values of the angles $\left(\theta_1^{(0)},\theta_1^{(0)}\right)$, and solve the minimum surface connecting $p_1$ and $p_2$, and compute the $Q_1$ and $Q_2$.
  \item To find the $(\theta_1,\,\theta_2)$ such that $Q_1 = Q_2 = 0$, we deduce the correction $\delta\theta_1,\,\delta\theta_2$ using the Newton-Raphson method as,
        \begin{equation}\label{eq:gradient}
          \left(
          \begin{array}{l}
              Q_1 \\
              Q_2
            \end{array}
          \right)
          + \left(
          \begin{array}{ll}
              \partial_{\theta_1} Q_1 & \partial_{\theta_2} Q_1 \\
              \partial_{\theta_1} Q_2 & \partial_{\theta_2} Q_2
            \end{array}
          \right)
          \left(
          \begin{array}{l}
              \delta \theta_1 \\
              \delta \theta_2
            \end{array}
          \right)
          =0.
        \end{equation}
        The Jacobian element can be approximated with $\partial_{\theta_i} Q_j \simeq \frac{Q_{j}(\theta_i+\delta\theta_i)-Q_j(\theta_i)}{\delta\theta_i}$, which requires solving the minimum surface at least three times.
  \item Solve the linear equation \eqref{eq:gradient}, and obtain the corrections $(\delta \theta_1,\delta \theta_2)$. Update $\theta_1,\,\theta_2$ with $(\theta_1,\theta_2) = (\theta_1,\theta_2) + (\delta \theta_1,\delta \theta_2)$.
  \item Iterate the above three steps, until $Q_1=0$ and $Q_2=0$ is satisfied within error bound. In this paper, we set the error bound as $10^{-6}$, where only solutions with $|Q_i| < 10^{-6}$ are accepted.
\end{enumerate}

A careful choice of the initial values $(\theta_1,\, \theta_2)$ is needed for the iterations to converge. The numerical reliability is guaranteed by the convergence of the results when setting different initial values\footnote{In principle, there may be local minimums of EoP, thus different initial values need to be assigned to check this. In this paper, we obtained the same results for different initial values, which shows that the area of the cross-section is globally convex in $(\theta_1,\, \theta_2)$.} or increasing the density of discretization (see \cite{Boyd:2001} for more technical details). Compared with the previous method, the current method is more advanced in the following aspects,

\begin{enumerate}
  \item The iteration is fast as long as we iterate a solution with a good initial value. A good strategy is to use a solution as the initial solution when solving a problem with parameters nearby.
  \item The solution is more precise compare with the previous method. In this paper, we can obtain results with $ |Q_i| \sim 10^{-7}$.
  \item It does not suffer from the coordinate singularity like the previous method, and hence the results are much more stable. Which means that it can obtain solutions in a larger range of parameters.
\end{enumerate}

Next, based on the above techniques, we explore the relationship between HEE, MI, EoP and phase transition, as well as the comparison between them.

\section{The Holographic entanglement entropy}\label{sec:hee}

\begin{figure}[]
  \centering
  \includegraphics[width =0.7\textwidth]{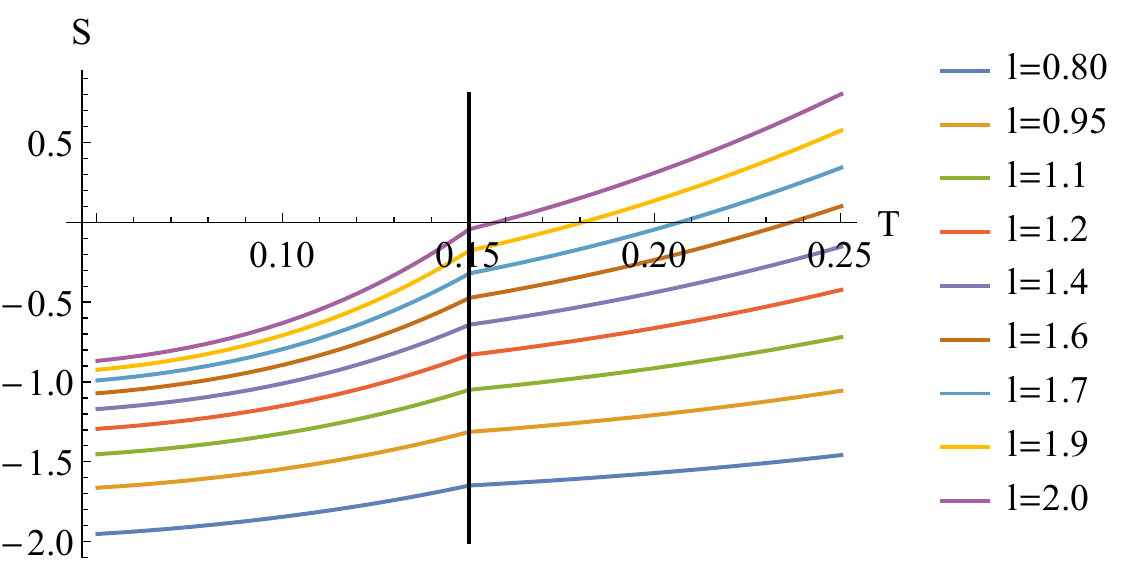}
  \caption{The HEE vs $T$ at different values of $l$ specified by the plot legends. The black vertical line labels the critical temperature.}
  \label{fig:heebehavior}
\end{figure}

\begin{figure}[]
  \centering
  \includegraphics[width =0.5\textwidth]{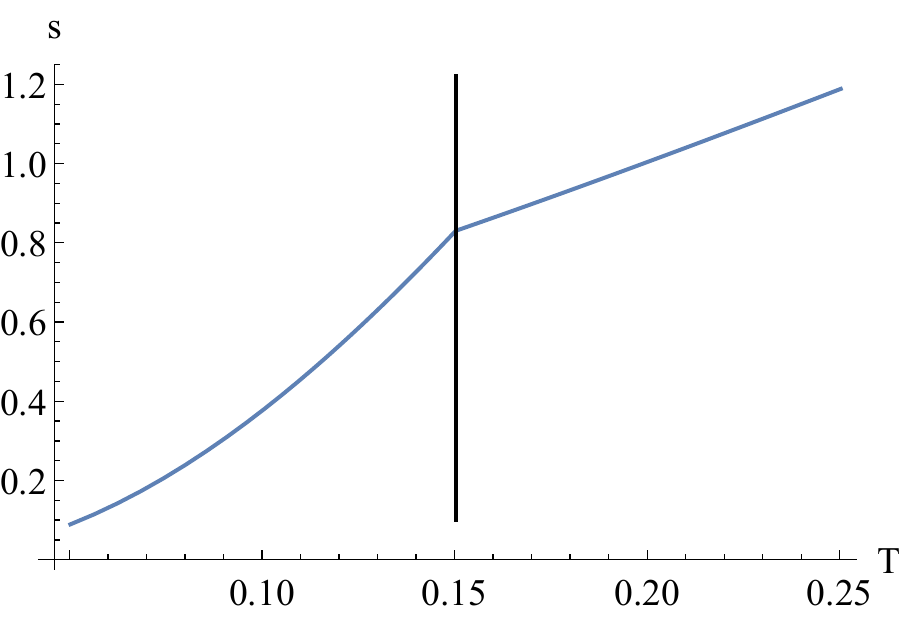}
  \caption{The entropy density $s$ vs $T$. The black vertical line labels the critical temperature.}
  \label{fig:entdensity}
\end{figure}

Fig. \ref{fig:heebehavior} shows the relationship between HEE and temperature in the critical region. As can be seen from Fig. \ref{fig:heebehavior}, HEE is continuous at the critical point, but its first derivative with respect to $T$ is discontinuous. In addition, HEE increases with increasing temperature in the critical region. These phenomena do not depend on the width $l$ of the infinite strip. It has been shown that the larger the width of the infinite band, the greater the contribution of thermodynamic entropy in HEE \cite{Ling:2016ibq}. Therefore the thermodynamic entropy, as a quantity which only depends on the near horizon geometry, should also diagnose the superconducting phase transition. This is certified in Fig. \ref{fig:entdensity}, where the thermal entropy density $s$ indeed show similar phenomena as the HEE.

The above phenomena show that both HEE and the thermal entropy are good diagnose of the thermal phase transition. This is as expected since the thermal phase transition is always accompanied by the emergence of condensation. Such condensation will introduce new degrees of freedom to the system, and hence radically change the thermal entropy properties, as well as the EE. Similar phenomena of the HEE over superconductivity phase transition have been obtained in \cite{Cai:2012sk,Cai:2012nm,Peng:2015yaa,Yao:2016ils}.

With the HEE, the MI is readily computed.

\section{The holographic mutual information}\label{sec:mi}

\begin{figure}[]
  \centering
  \includegraphics[width =0.7\textwidth]{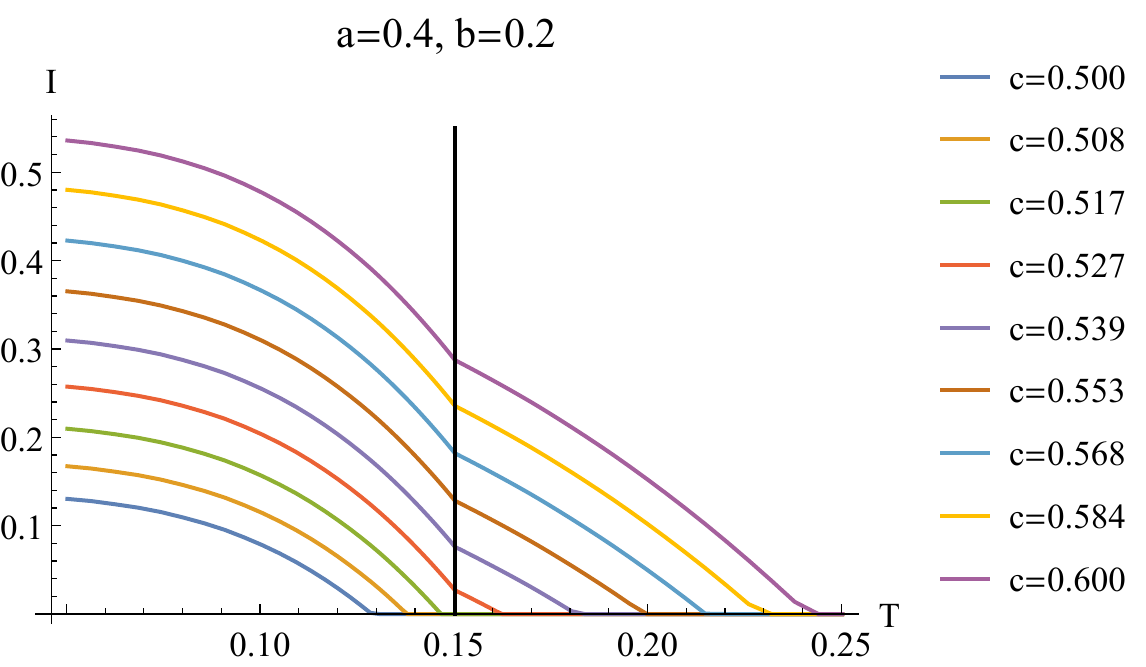}
  \includegraphics[width =0.7\textwidth]{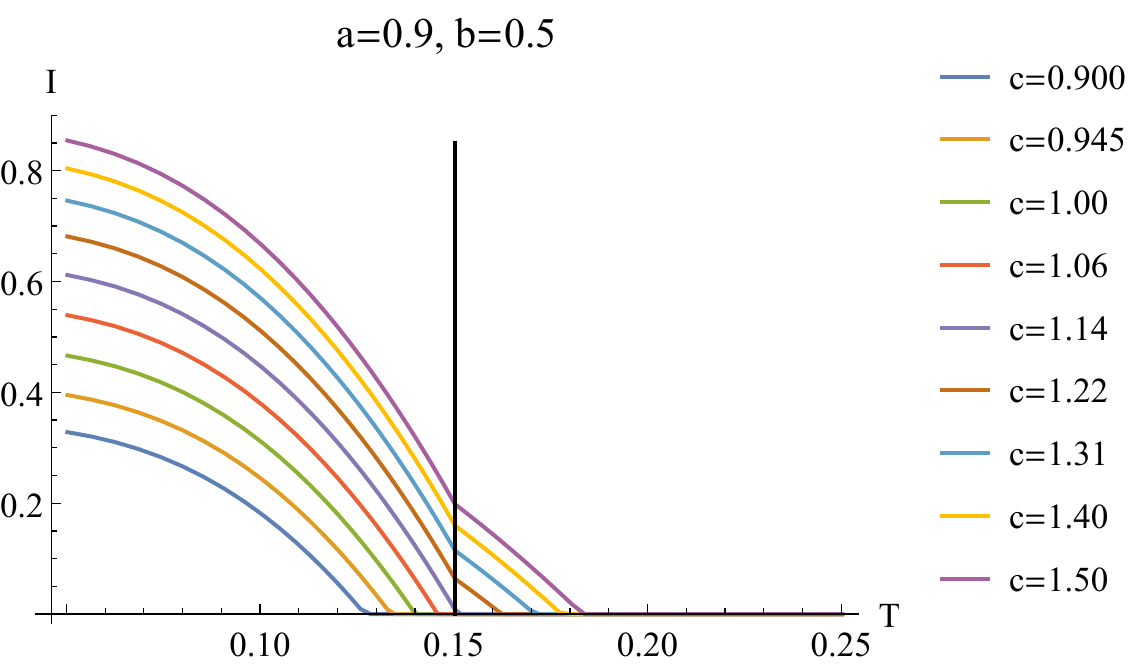}
  \caption{The MI vs $T$. The black vertical line labels the critical temperature.}
  \label{fig:mibehavior}
\end{figure}

The MI, originated from the EE, should also reflect the thermal phase transition. Fig. \ref{fig:mibehavior} shows the relationship between MI and temperature in the critical region. First, MI decreases with increasing temperature, which is opposite to the relationship between HEE and temperature. Secondly, similar to HEE, MI can diagnose phase transitions indeed. MI is continuous at the critical point, but its first derivative with temperature is discontinuous. In addition, when the temperature increases, the MI may decrease to zero, which is called the disentangling phase transition. This can be understood as that thermal effects may destroy the quantum entanglement. Also, the system disentangles more easily for smaller values of $c$ when fixing the $a,\,b$.

\begin{figure}[]
  \centering
  \includegraphics[width =0.75\textwidth]{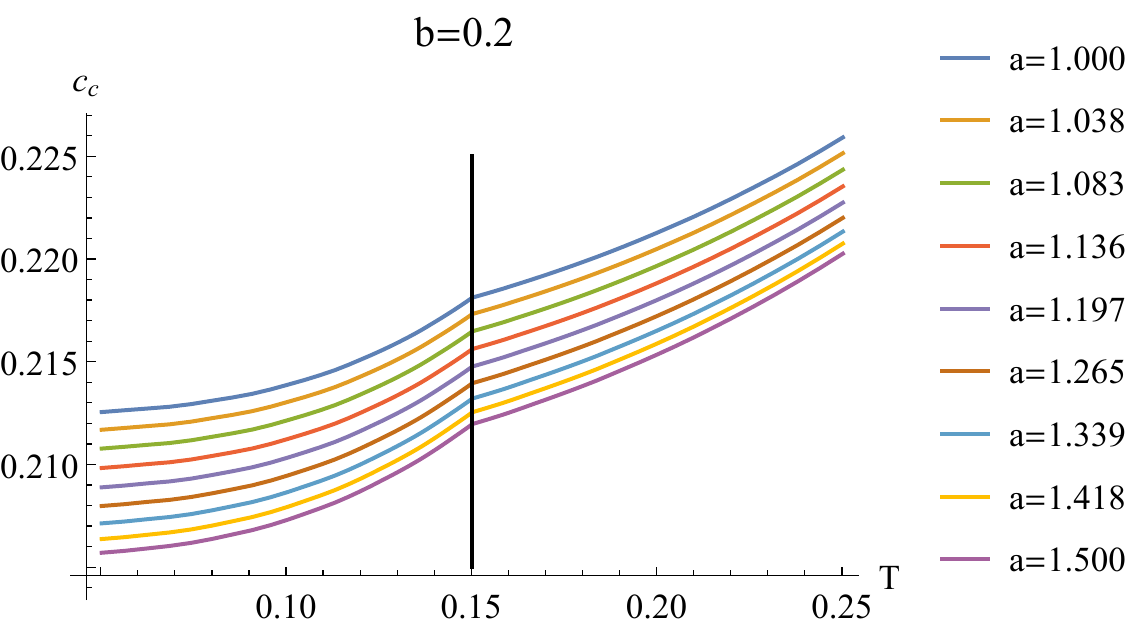}
  \caption{critical $c$}
  \label{fig:criticalc}
\end{figure}

An interesting quantity related to the disentangling phase transition is the critical size of the configurations. We demonstrate the critical $c$ (labeled as $c_c$) in Fig. \ref{fig:criticalc}, in which we see that the critical $c$ increases with the increasing temperature. This is in accordance with the phenomena observed in Fig. \ref{fig:mibehavior}.

Another notable feature is that MI is always convex in the critical region. In contrast, HEE is concave. The opposite behavior of MI and HEE is essentially due to their association on definition. EoP is different from MI and HEE by definition, thus its behavior in the critical region is worth exploring.

\section{The holographic entanglement of purification}\label{sec:eop_phenomena}

In this section, we first explore the EoP behavior in the critical region to reveal the relationship between EoP and phase transition. Subsequently, we will verify that the EoP in this paper still satisfies some important inequalities.

\subsection{EoP and thermal phase transition}

\begin{figure}[]
  \centering
  \includegraphics[width =0.48\textwidth]{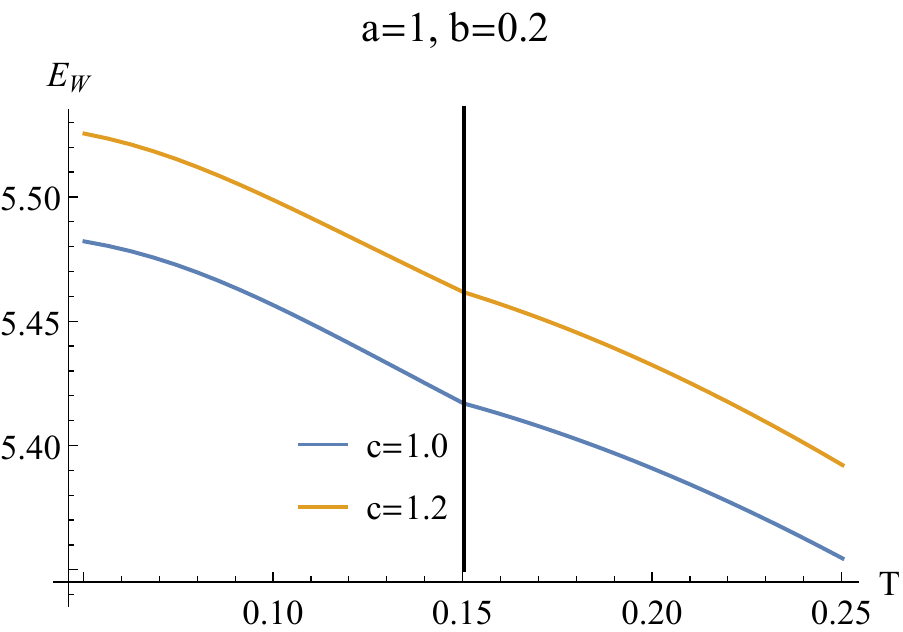}
  \includegraphics[width =0.48\textwidth]{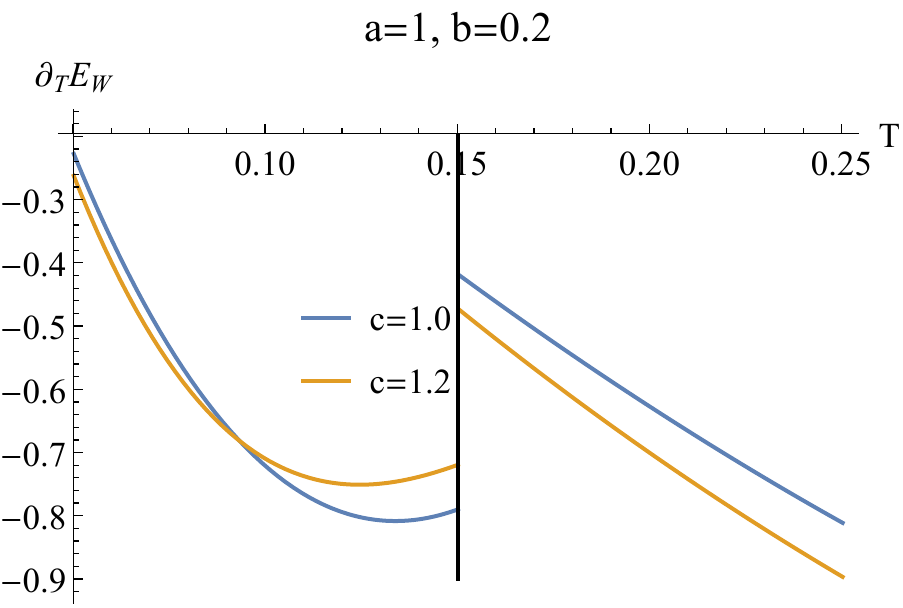}
  \caption{The EoP $E_W$ and the first order temperature derivative with respect to $E_W$ vs the temperature. These two plots are obtained at $(a,b)=(1,0.2)$ at different values of $c$ specified by the plot legends. Here, the $E_W$ is convex.}
  \label{fig:asymeopcase2}
\end{figure}

\begin{figure}[]
  \centering
  \includegraphics[width =0.48\textwidth]{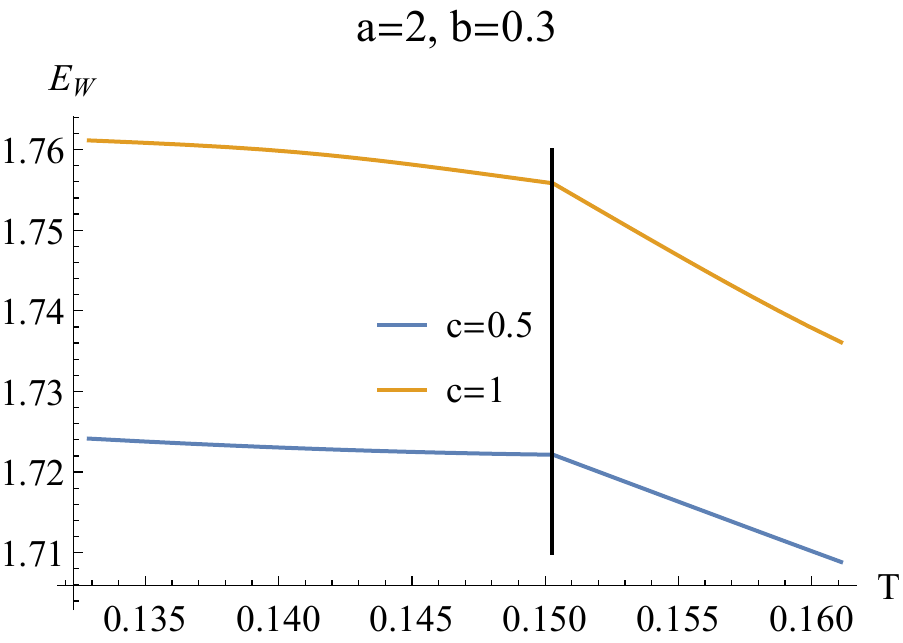}
  \includegraphics[width =0.48\textwidth]{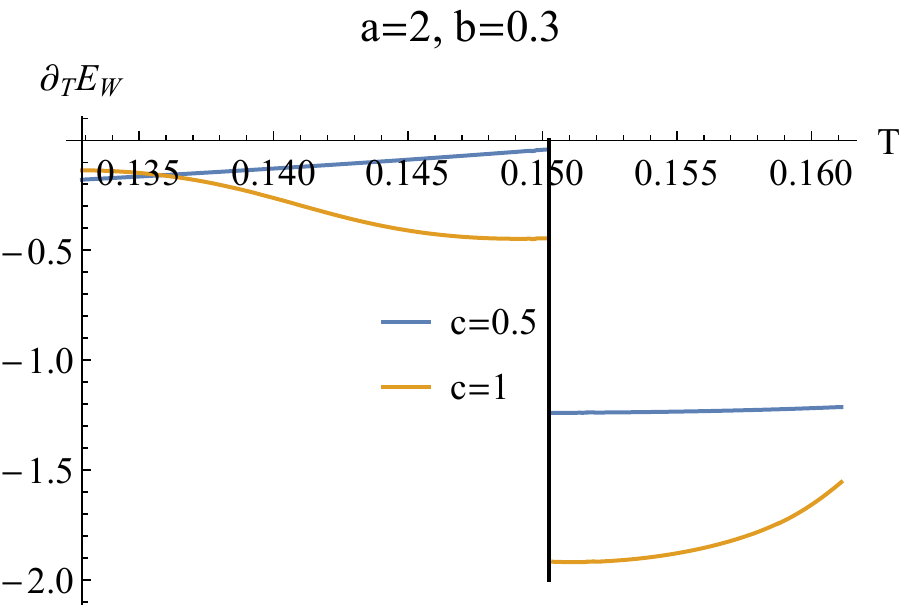}
  \caption{The EoP $E_W$ and the first order temperature derivative with respect to $E_W$ vs the temperature. These two plots are obtained at $(a,b)=(2,0.3)$ at different values of $c$ specified by the plot legends. Here, the $E_W$ is concave.}
  \label{fig:asymeopcase1}
\end{figure}

Similar to HEE and MI, EoP also shows obvious unsmoothness at the critical point. As shown in Fig. \ref{fig:asymeopcase2}, EoP is continuous, but its first derivative is discontinuous at the critical point. Moreover, EoP decreases with increasing temperature, which is consistent with MI. In addition, it can exhibit convex behavior in the critical region like MI. Intriguingly, EoP can also exhibit concave behavior (as shown in Fig. \ref{fig:asymeopcase1}). This is a key difference between EoP and MI.

Whether the EoP is convex or concave depends on specific configurations.
By comparing the Fig. \ref{fig:asymeopcase2} and Fig. \ref{fig:asymeopcase1}, we can find that when the configuration is relatively small and the the minimum cross-section is far away from the horizon, the EoP exhibits a convex behavior similar to that of MI. However, when the configuration is relatively large, where the minimum cross section is close to the event horizon of the black hole, the EoP will show a concave behavior.

The difference between EoP and MI shows that EoP, as a measure of mixed state entanglement, characterizes different information of quantum system. Moreover, the configuration-dependent properties of EoP show that EoP exhibits more abundant phenomena than MI, which may reveal the properties of quantum entanglement more comprehensively.

\begin{figure}
  \centering
  \includegraphics[width=0.6\textwidth]{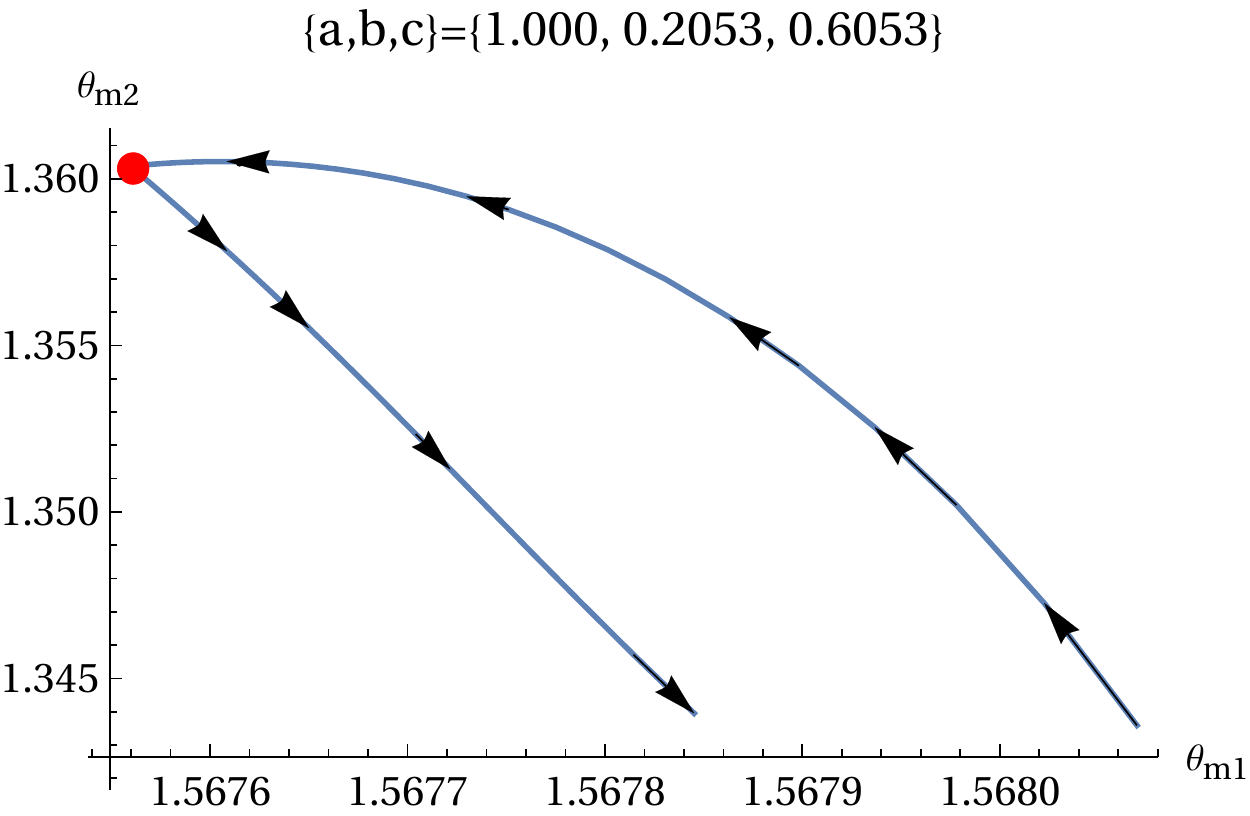}
  \caption{The endpoints of the corresponding EoP for different values of temperature. Along the arrow direction, the temperature increases, and the turning point (red point) exactly matches the critical temperature.}
  \label{fig:anglebehavior}
\end{figure}

Another interesting phenomenon is the angle (endpoint) behavior in the critical region. From Fig. \ref{fig:anglebehavior} we find that the angle parameter can work as a good diagnose of the phase transition by showing a rapid turn in the space of the $(\theta_1,\theta_2)$. Also, the typical change of $\theta$ when variating the $T$ is of order $10^{-4}$, which has been well captured by our numerics that is precise up to $10^{-7}$. When changing $T$, the $\theta_{m1}$ changes very slowly, while the $\theta_{m2}$ changes relatively more rapidly. This is as expected, since the point $p_1$ parametrized by $\theta_{m1}$ locates at the regions near the boundary, where the major contribution to EoP lies in. Therefore, it will change less than that of the $p_2$ (parametrized by $\theta_{m2}$) region, that is relatively far away from the boundary.

\subsection{Inequalities of EoP}

The EoP satisfies several important inequalities, that the correct holographic EoP expression must satisfy.

\begin{figure}
  \centering
  \includegraphics[width=0.7\textwidth]{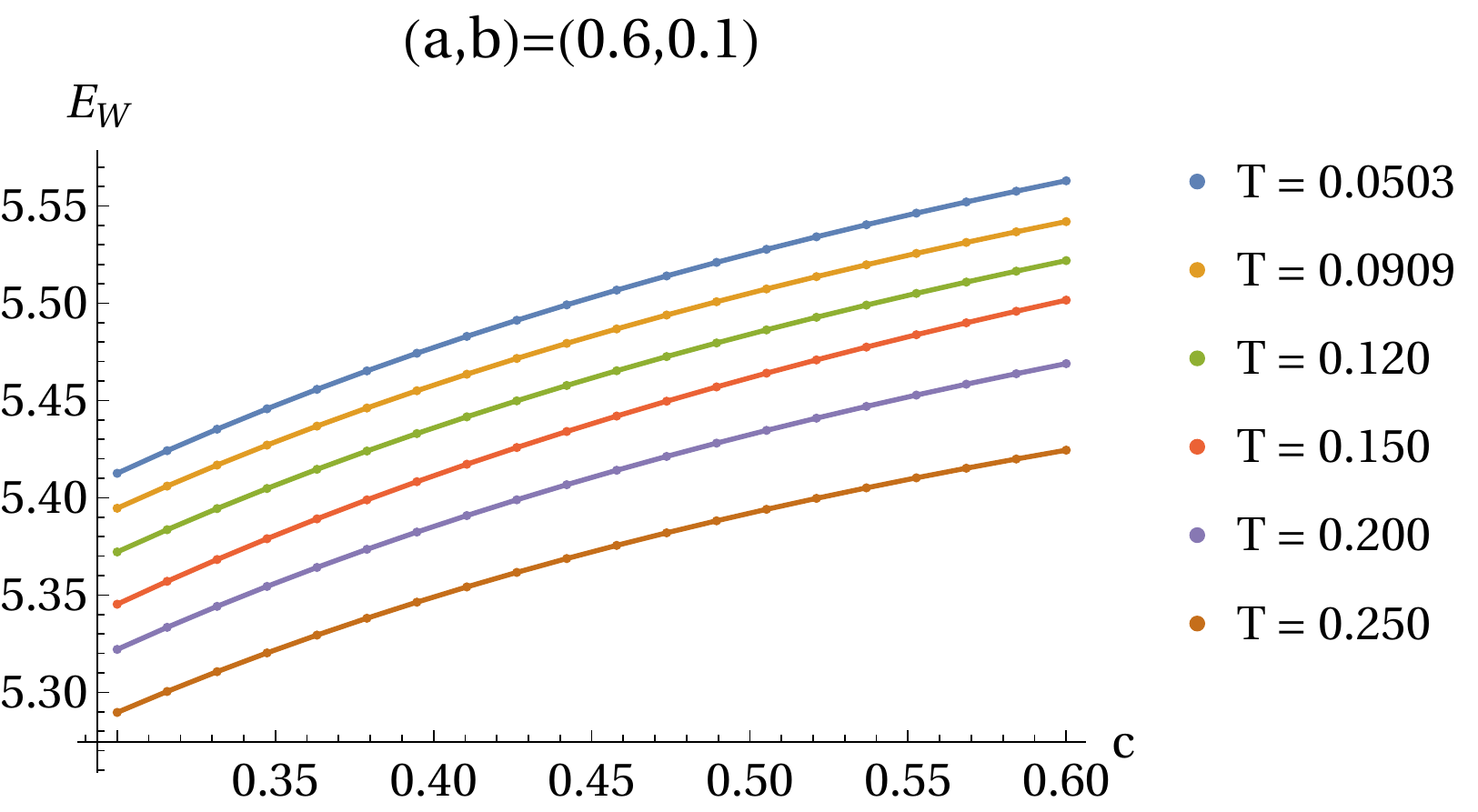}
  \caption{EoP vs $c$. Apparently, the EoP decreases with the temperature. Also, EoP increases as the $c$ increases, this is one of the inequalities that EoP has to satisfy.}
  \label{fig:eopvsc}
\end{figure}

The first inequality is
\begin{equation}
  \label{eq:ineq_1}
  E _ { W } \left( \rho _ { A ( B C ) } \right) \geqslant E _ { W } \left( \rho _ { A B } \right)
\end{equation}
which has bene shown with the entanglement wedge nesting property \cite{Takayanagi:2017knl}. The inequality \eqref{eq:ineq_1} can be translated into
\begin{equation}\label{eq:prove1}
  E_{W}\left(a,b,c+\delta c\right)\geqslant
  E_{W}\left(a,b,c\right) \quad \text{with} \quad\delta c\geqslant 0.
\end{equation}
This is readily seen in Fig. \ref{fig:eopvsc}, where we can find that $E_W$ indeed increases with increasing $c$ at fixed values of $a,\,b$ and $T$.

\begin{figure}
  \centering
  \includegraphics[width =0.6\textwidth]{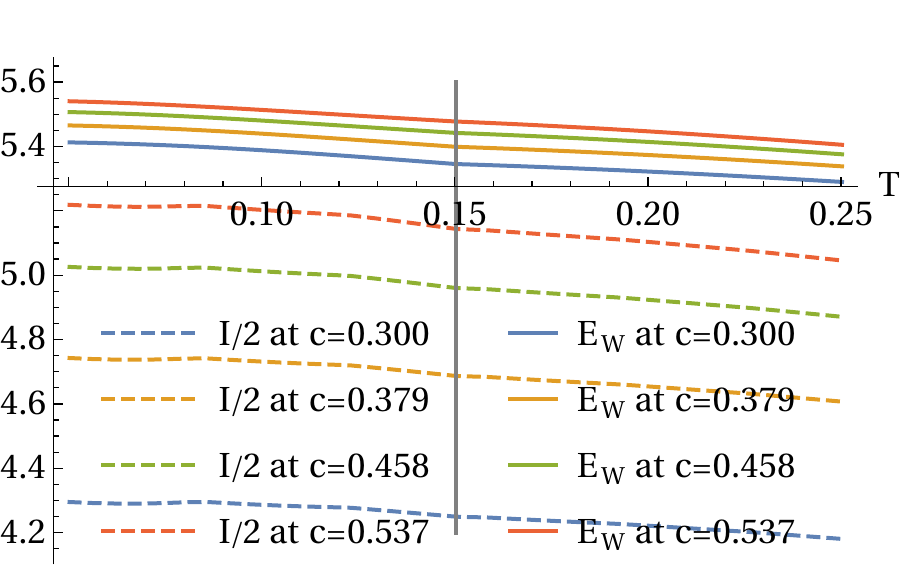}
  \caption{EoP vs $I/2$. The solid lines and the dashed lines are the EoP and one half of the MI at different values of $c$, respectively. At a fixed value of $c$, the solid line and the dashed line are of the same color. It is readily seen that EoP is always greater than one half of the MI. In this plot we fix $(a,b)=(0.6,0.1)$.}
  \label{fig:eopvshalfmi}
\end{figure}

The second inequality is,
\begin{equation}
  \label{eq:ineq_2}
  E _ { W } \left( \rho _ { A C } \right) \geqslant \frac { 1 } { 2 } I ( A , C ),
\end{equation}
which states that $E_W$ of any configuration is greater than half of the MI. This is shown in Fig. \ref{fig:eopvshalfmi}, where the data of the solid curves (EoP) of a certain color is always larger than that of the dotted curves (EoP) of the same color.

The above two phenomena show that the EoP in this paper does satisfy the important inequalities. These results once again enhance the reliability of holographic EoP prescription.

\section{Discussion}\label{sec:discuss}

We have investigated the HEE, MI and EoP for general strip configurations in superconductivity phase transition model. We find that HEE, MI and EoP can all diagnose the superconducting phase transition. They are continuous at the critical point, but their first derivative with respect to temperature is discontinuous. In addition, as a measure of entanglement of mixed states, MI exhibits the opposite behavior to HEE. Specifically, MI decreases with increasing temperature and exhibits a convex behavior, while HEE increases with increasing temperature and exhibits a concave behavior. These results do not depend on the specific configuration. Moreover, as a new measure of mixed state entanglement, EoP can exhibit either the same or the opposite behavior as MI, depending on the size of the specific configuration. These results show that EoP can not only describe the phase transition, but also capture more abundant information than HEE and MI.

Thermal phase transition is usually accompanied by the emergence of order parameter \cite{Cai:2015cya}, which is the main reason why HEE, MI and EoP can diagnose it. However, not all phase transitions occur with the emergence of order parameters. Quantum phase transition occurs at zero temperature when changing system parameters. There are certain quantum phase transitions in which the order parameter is absent. Therefore, the characterization of these quantum phases becomes an important topic. Metal-insulator transition, as one of the most well-known quantum phase transition, was found intimately related to the EE \cite{Ling:2015dma,Ling:2016wyr,Ling:2016dck}. However, EE cannot completely exclude the contribution of thermal entropy. As a new mixed state entanglement measure independent of the EE, we can expect that EoP may play an important role in quantum phase transition. This is the direction of our future efforts.

Another major advance of this paper is to provide an upgraded version of EoP algorithms. Using these algorithms, the calculation of EoP can be more stable and reliable, which can pave the way for further study of EoP. For example, the properties of EoP in Born-Infeld system, massive gravity and Lovelock gravity theory are all worth exploring. These studies will lay a foundation for a more comprehensive understanding of the properties of mixed state entanglement in more general holographic models.

\section*{Acknowledgments}

We are very grateful to Long Chen, Wu-Zhong Guo, Peng-Xu Jiang, Zhuo-Yu Xian for helpful discussions and suggestions. Peng Liu would like to thank Yun-Ha Zha for her kind encouragement during this work. This work is supported by the Natural Science Foundation of China under Grant No. 11575195, 11875053, 11805083, 11847055, 11905083, 11775036 and Fok Ying Tung Education Foundation under Grant No. 171006. Jian-Pin Wu is also supported by Top Talent Support Program from Yangzhou University.

\end{document}